\documentclass[pra,twocolumn,amsmath,amssymb,superscriptaddress]{revtex4-1}
\usepackage{epsfig, graphicx,graphics,amsmath,amssymb,float}
\usepackage[T1]{fontenc}
\usepackage[latin9]{inputenc}
\usepackage{amsmath}
\usepackage{amssymb}
\usepackage{xcolor}
\usepackage{amscd}
\usepackage{bm}
\usepackage{bbold}
\usepackage{psfrag}

\usepackage{bbm}

\usepackage[caption=false]{subfig}
\usepackage{subfig}
\usepackage{color}
\usepackage[bookmarks=true,colorlinks,linkcolor=blue,urlcolor=blue,citecolor=blue]{hyperref}

\DeclareMathOperator{\sgn}{sgn} 

\newcommand{\be}{\begin{equation}}
\newcommand{\ee}{\end{equation}}
\newcommand{\bea}{\begin{eqnarray}}
\newcommand{\eea}{\end{eqnarray}}

\newcommand{\mc}{\mathcal}
\newcommand{\mb}{\mathbf}

\begin{document}

\title{Quench dynamics of the Kondo effect:\\ transport across an impurity coupled to interacting wires}

\author{Moallison F. Cavalcante}
\affiliation{Departamento de F\'isica, Universidade Federal de Minas Gerais, C. P. 702, 30123-970, Belo Horizonte, MG, Brazil}
\affiliation{Universit\'e Paris-Saclay, CNRS, Laboratoire de Physique des Solides, 91405, Orsay, France}
\author{Rodrigo G. Pereira}
\affiliation{International Institute of Physics and Departamento de F\'isica Te\'orica e Experimental, Universidade Federal do Rio Grande do Norte, 59072-970 Natal-RN, Brazil}
\author{Maria C. O. Aguiar}
\affiliation{Departamento de F\'isica, Universidade Federal de Minas Gerais, C. P. 702, 30123-970, Belo Horizonte, MG, Brazil}
\date{\today}

\begin{abstract}
We study   the real-time dynamics of the Kondo effect after a quantum quench in which a  magnetic impurity is   coupled to two metallic Hubbard chains. Using an effective field theory approach, we find that for noninteracting electrons  the charge current across  the impurity is given by a scaling function that involves the Kondo time.  In the interacting case, we show   that the Kondo time  decreases with the strength of the repulsive interaction and the time dependence of the  current reveals signatures of the Kondo effect in a Luttinger liquid.  In  addition, we verify that the relaxation of the impurity magnetization does not exhibit universal  scaling behavior in the perturbative regime below the Kondo time. Our results highlight the role of nonequilibrium  dynamics as a valuable tool in the study of  quantum impurities in interacting systems.

\end{abstract}

\pacs{pacs}

\maketitle

\section{Introduction}
\label{sec1}
Quantum impurity problems in low-dimensional systems have been of great importance to the understanding of   many-body   systems for a long time \cite{Affleck2010}. The interaction between the impurity and the bulk degrees of freedom can be approached theoretically using well-established  analytical approaches based on exact solutions \cite{Wiegmann1980,Andrei1983} and  boundary conformal field theory \cite{Cardy1984,CFT}  and numerical  methods such as the numerical renormalization group \cite{Bulla2008} and the  density matrix renormalization group (DMRG) \cite{DMRG}. At the same time,  a steady  development of  experimental    techniques to control and probe synthetic  quantum matter in quantum dots \cite{Eq09,dot2} and ultracold atoms \cite{ultracold2, nishida2013,riegger2018localized,Kanasz-Nagy2018}  opens the possibility of testing many theoretical scenarios.

A currently active   area of research  is the study of the real-time dynamics of quantum many-body systems driven out of equilibrium \cite{Eisert2015,Mitra}. A simple way of simulating   the non-equilibrium dynamics in closed systems is by means of quantum quench protocols~\cite{Calabrese,Cazalilla,Calabrese2,Mitra}. Consider a system described by a Hamiltonian $H(g)$, where $g$ stands for  a parameter or a set of parameters, and suppose that the system  is initially prepared in the ground state $|\Psi_0\rangle$ of $H(g)$. A quantum quench is defined by a sudden (much faster than any other characteristic internal time scale) change $H \to H' = H(g')$, followed by the unitary evolution of the system under $H'$.

In this work, we investigate  the formation of the Kondo effect  \cite{Kondobook} in the real-time dynamics   following a quantum   quench. In electronic systems, the Kondo effect  arises when the spin of a localized magnetic impurity couples to        conduction electrons via an antiferromagnetic exchange interaction with Kondo coupling $J_K>0$. The hallmark of this effect is the emergence of an energy scale, $k_BT_K$, where $k_B$ is the Boltzmann constant and $T_K$ the Kondo temperature, which marks a crossover from    weak coupling at  temperatures $T\gg T_K$  to    strong coupling at $T\ll T_K$. The crossover can be detected  in various observables that behave as scaling functions of $T/T_K$. For instance,  at high temperatures the impurity magnetic susceptibility exhibits  a logarithmic scaling, $\chi_{\rm imp}\sim \ln (T_K/T)$, characteristic of the perturbative renormalization   of the effective Kondo coupling. At $T\lesssim T_K$, perturbation theory in the Kondo coupling breaks down, and the low-temperature regime $T\ll T_K$ is described by    the localized spin forming a singlet state with a conduction electron. In this regime,  the impurity  susceptibility shows a $(T/T_K)^2$ dependence governed by irrelevant perturbations to the strong-coupling fixed point \cite{Nozieres1974}.  In   quantum wires with finite  length $L$, the crossover can occur at zero temperature as a function of the ratio $L/\xi_K$,  where $\xi_K=\hbar v_F/(k_BT_K)$ (with $v_F$   the Fermi velocity)  is the size of the Kondo cloud that surrounds and screens the localized spin \cite{Barzykin1996,Simon2001,Simon2003,Eq07,Pereira2008,affleck2010kondo}. The Kondo  cloud was recently observed in a mesoscopic device~\cite{Borzenets2020}. By analogy and dimensional analysis, one can argue for the existence of a Kondo time   $t_K=\hbar/(k_BT_K)$. In fact, the latter shows up  in time-dependent response functions \cite{1kondo} and can be interpreted as the time scale for the formation of the Kondo cloud after the Kondo coupling is suddenly switched on.  The analogy with the equilibrium Kondo effect has motivated the search for universal scaling behavior in the time evolution after    quenches in  quantum impurity models  \cite{TD-NRG,3kondo,Lobaskin2005,Hackl2009,Pletyukhov2010,4kondo,Andergassen2011,Medvedyeva2013,Vasseur2013,Lechtenberg2014,Kennes2014,Kennes2014,Nuss2015,Ghosh2015,Antipov2016,Costi2017,Krivenko2019,Goto2019}.

\begin{figure}[t]
    \centering
    \includegraphics[scale=0.35]{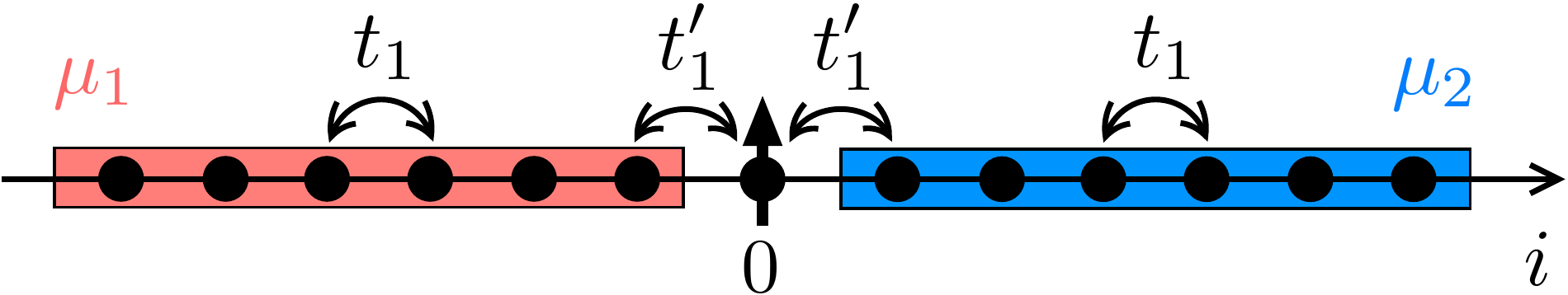}
    \caption{Schematic setup for a quantum dot coupled to two semi-infinite chains. The hopping parameter in both chains is $t_1$. The  quantum dot state  at $i=0$ has  energy $\epsilon_d<0$ and on-site electron-electron interaction $U_d>0$. In the quench protocol, the single electron in the dot is initially polarized and the left and right chains are held at chemical potentials $\mu_1$ and $\mu_2$, respectively. At time $t=0$,   the dot is coupled to the chains with  hopping  parameter $t_1'$. After that,  a charge current starts flowing  across  the dot,  with a time dependence that reveals signatures   of the Kondo effect.  }
    \label{system}
\end{figure}

Our goal  is to observe the emergence of the Kondo time scale within an analytical approach for a quench protocol that probes charge transport across a magnetic impurity. Similar protocols have been studied in the anisotropic Kondo model  with noninteracting leads \cite{Pletyukhov2010,Hackl2009} and in junctions of Luttinger liquids far from equilibrium without the localized spin \cite{Schiro2015}. Here we   study the isotropic Kondo model   including interactions in the leads. The schematic  setup is shown in Fig. \ref{system}. We  consider two electronic chains held at different chemical potentials and   coupled to a singly occupied  quantum dot that acts as an $S=1/2$ magnetic impurity. In the static problem in the linear response regime, this setup  reveals signatures of the Kondo effect as the conductance across the dot scales with $T/T_K$,  approaching the maximum value $2e^2/h$ for $T\to 0$ in the particle-hole symmetric case \cite{Pustilnik, Affleck2010}. Here we shall look for scaling behavior as a function of the time ratio $t/t_K$ after the chains are suddenly coupled to the impurity spin prepared in a polarized state. We consider both regimes of $t\ll t_K$ and $t\gg t_K$, governed by the weak- and strong-coupling fixed points of the Kondo model, respectively. Our results indicate that the charge current  in the post-quench dynamics can be described by a scaling function of $t/t_K$ in the case of noninteracting electrons in the leads. We then turn to  interacting chains described by the Hubbard model and discuss how Luttinger liquid effects modify the exponents in the time dependence of the current as the system approaches the steady state. In this process, we also generalize   previous results  \cite{Helena} for the real-time decay of the impurity magnetization, and show that the latter is not a universal function in the regime $t\ll t_K$.  

This paper is organized as follows. In Sec. \ref{sec2} we briefly review the derivation of the Kondo model for a  quantum dot embedded between two tight-binding  chains. We also introduce the quench protocol and discuss the time scales involved in the problem. In Sec. \ref{sec4} we consider the current dynamics in the case of noninteracting electrons in the leads, identifying the scaling behavior as a function of $t/t_K$ in both weak- and strong-coupling limits. In Sec. \ref{sec5} we discuss the effects of electron-electron interactions in the chains. Section \ref{magneSec} is devoted  to the time dependence of the impurity magnetization. Our conclusions are summarized in Sec.\ref{concluSec}. Finally, Appendix \ref{ap1} contains the main  bosonization  formulas, and   Appendix \ref{ap2} focuses on the  three-point  function used in the perturbative calculations. Hereafter we set $\hbar = k_B = 1$.

\section{Model and Quench Protocol} 
\label{sec2}

We investigate the post-quench dynamics of a quantum dot  coupled to  two semi-infinite chains, see  Fig. \ref{system}. The system is described by the time-dependent Hamiltonian
\begin{equation}
    H(t) = H_0 +H_U+ H_d + \theta(t)H_{\text{coup}}.
    \label{begin}
\end{equation}
Here $H_0=\sum_{\ell=1,2}H_{\ell}$ is the Hamiltonian for decoupled chains  with \bea
     H_1 &=& -t_1\sum_{i\le -2}(c^{\dagger}_{i}c^{\phantom\dagger}_{i+1} + \text{h.c.}),
     \label{chain1}\\
     H_2 &=& -t_1\sum_{i\ge 1}(c^{\dagger}_{i}c^{\phantom\dagger}_{i+1} + \text{h.c.}) ,
     \label{chain2}
\eea
where $t_1$ is the hopping parameter in the chains and $c^{\dagger}_i=(c^{\dagger}_{i\uparrow},c^{\dagger}_{i\downarrow})$, with  $c^\dagger_{i\sigma}$ the creation operator for an electron with spin $\sigma$ at site $i$. The corresponding number operator  is $n_{i\sigma}=c^{\dagger}_{i\sigma}c^{\phantom\dagger}_{i\sigma}$. The second term in  Eq. (\ref{begin}) accounts for electron-electron interactions in the chains:\be
H_{U}=U\sum_{i\neq0}n_{i\uparrow}n_{i\downarrow},
\ee
where $U\geq 0$ is the strength of the   on-site repulsive interaction. The   Hamiltonian for the dot (denoted as site $i=0$) is given by
\begin{equation}
    H_d = \epsilon_dn_0 + U_dn_{0\uparrow}n_{0\downarrow},
\end{equation}
where  $n_0 = n_{0\uparrow}+n_{0\downarrow}$, $\epsilon_d$ is the energy of the localized state with respect to the Fermi level in the chains, and $U_d$ is the local interaction. To favor a local moment at the quantum dot, we consider  $\epsilon_d<0$ and $U_d>0$.  The coupling term reads
\begin{equation}
    H_{\text{coup}} = -t'_1\ (c^{\dagger}_{-1 } +  c^{\dagger}_{1 })c^{\phantom\dagger}_{0 } +\text{h.c.},
\end{equation}
where $t'_1$ is the hybridization between the dot and the end sites of  each chain. 

The Kondo regime corresponds to  $t'_1\ll -\epsilon_d, U_d$. In this case, we apply a Schrieffer-Wolff transformation to derive the effective Hamiltonian in the low-energy subspace with a single electron at the dot  \cite{Kondobook}. 
  We obtain
\begin{equation}
    H_{\rm eff} (t)= H_0+H_U+ \theta(t)(H_K+H_W),
    \label{H1}
\end{equation}
where
\begin{eqnarray}
H_{K}&=&J_K\mathbf S_0\cdot  (c^\dagger_{-1}+c^\dagger_{1} )\frac{\boldsymbol\sigma}{2} (c^{\phantom\dagger}_{-1}+c^{\phantom\dagger}_{1} ),\nonumber\\
H_W&=&W  (c^\dagger_{-1}+c^\dagger_{1} ) (c^{\phantom\dagger}_{-1}+c^{\phantom\dagger}_{1} ).\label{HKondo}
\end{eqnarray}
Here $H_K$ describes the Kondo interaction between conduction electrons in the symmetric channel and  the impurity spin 
  $\mathbf S_0=c^\dagger_{0}(\boldsymbol\sigma/2)c^{\phantom\dagger}_{0}$,  with   Kondo coupling  \begin{equation}
    J_K=2t'^2_1\left(\frac1{-\epsilon_d}+\frac{1}{U_d+\epsilon_d}\right).\label{JK}
\end{equation}
The  strength of the   potential scattering term $H_W$ is
\begin{equation}
    W=\frac{t'^2_1}{2}\left(\frac1{-\epsilon_d}-\frac{1}{U_d+\epsilon_d}\right).
\end{equation}
Note that $W$ vanishes in the   particle-hole symmetric case $\epsilon_d=-U_d/2$.

We now focus on  the model for noninteracting chains, $U=0$, and analyze it in the continuum limit. We will discuss the interacting case in Sec. \ref{sec5}. We replace $c_{j\sigma}$ by a fermionic field operator $c_{\ell,\sigma}(x)$, where $c_{1,\sigma}(x)$ is defined for $x<0$ and $c_{2,\sigma}(x)$ for $x>0$. To describe the low-energy modes in the chains, we expand  $c_{\ell,\sigma}(x)$ in terms of right (R) and left (L) movers: 
\begin{equation}
c_{\ell,\sigma}(x)=e^{ik_{F}x}\psi_{R,\ell,\sigma}(x) + e^{-ik_{F}x}\psi_{L,\ell,\sigma}(x),
\label{RL}
\end{equation}
where $k_F$ is the Fermi momentum, assumed to be the same for both chains at equilibrium. Particle-hole symmetry imposes  half filling, $k_F=\pi/2$.  For $t'_1=0$, the open boundary condition $c_{\ell,\sigma}(0)=0$ can be cast as a constraint on  the chiral fermionic modes  in each wire: \be
\psi_{L,\ell,\sigma}(x)=-\psi_{R,\ell,\sigma}(-x).
\label{constr}
\ee
The above constraint allows us to work with  a single chiral mode redefined in the  domain $x\in \mathbbm R$ \cite{Fabrizio1995}:\bea
\psi^\dagger_{1}(x)&\equiv&(\psi^\dagger_{L,1,\uparrow}(-x),\psi^\dagger_{L,1,\downarrow}(-x)),\nonumber\\
\psi^\dagger_{2}(x)&\equiv&(\psi^\dagger_{R,2,\uparrow}(x),\psi^\dagger_{R,2,\downarrow}(x)).
\eea
In the continuum limit, the noninteracting Hamiltonian in the leads can be written as 
\begin{equation}
    H_0=v_F \sum_\ell  \int dx\,\psi^\dagger_{\ell}(-i\partial_x)\psi^{\phantom\dagger}_{\ell},    \label{Hwire}
\end{equation}
with   Fermi velocity $v_F=2t_1\sin k_F$. For small $t'_1$, we can treat the local interactions as perturbations to the weak-coupling fixed point $J_K=W=0$. In terms of the fermionic fields, we have
\begin{eqnarray}
  H_K&=&\pi v_F\lambda_K\mathbf S_0 \cdot\sum_{\ell,\ell'}:\psi^{\dagger}_{\ell}(0)\frac{\boldsymbol\sigma}{2}\psi^{\phantom\dagger}_{\ell'}(0): ,\label{JK1} \\
H_W  &=& W'\sum_{\ell,\ell'}:\psi^{\dagger}_{\ell}(0)\psi^{\phantom\dagger}_{\ell'}(0):,   \label{HW}
\end{eqnarray}
where  $ \lambda_K=4J_K\sin^2k_F/(\pi v_F)$ is the dimensionless Kondo coupling,   $W' = 4W\sin^2k_F$, and $: \,:$ denotes normal ordering. 

In the static problem,   the effective Kondo coupling  at energy scale $\Lambda$ obeys the renormalization group (RG) equation \cite{CFT, Kondobook}
\begin{equation}
    \frac{d}{dl}\lambda_K = \lambda^2_K + \mc O(\lambda^3_K),
    \label{RG1}
\end{equation}
where $l=\ln(\Lambda_0/\Lambda)$ with $\Lambda_0\sim t_1$ the bare cutoff. In contrast, the potential scattering parameter $W'$ is strictly marginal. As a result, the low-energy physics is dominated by the flow of $\lambda_K$ to strong coupling, in the form 
\begin{equation}
    \lambda^{\rm eff}_K(l) \approx \frac{\lambda_0}{1-\lambda_0l}, \label{efflambda}
\end{equation}
where $\lambda_0=\lambda^{\rm eff}_K(0)$ is the bare coupling constant.  In the lattice picture for the strong-coupling fixed point $\lambda_K \to \infty$, the impurity forms a singlet with an electron  in the symmetric orbital  associated with  $(c^{\dagger}_{1}+c^{\dagger}_{-1})/\sqrt{2}$ \cite{Simon2001,Simon2003}. At low energies, this symmetric orbital   is blocked by a binding energy of order $ T_K\sim \Lambda_0 e^{-1/\lambda_0}\ll \Lambda_0$. This effect changes the boundary conditions for the symmetric channel, but electrons can move freely between  the chains   through the anti-symmetric orbital associated with $(c^{\dagger}_{1}-c^{\dagger}_{-1})\sqrt{2}$. For $W'=0$, the strong-coupling fixed point is characterized by the ideal conductance $G_0=e^2/\pi$ (in unit of $\hbar=1$).  More generally, the conductance is lowered by the potential scattering term allowed in the case of broken particle-hole symmetry \cite{Simon2001}.

To study the post-quench dynamics of the model, we consider that for times $t<0$ the system is prepared  in the state\be
|\Psi_0\rangle=|\psi_0\rangle_1\otimes \left|\uparrow \right\rangle \otimes|\psi_0\rangle_2 , \label{Psi0}
\ee
where $|\psi_0\rangle_\ell$ for $\ell=1,2$ denote the ground states of the disconnected  chains and  $\left|\uparrow \right\rangle$ is the spin-polarized state of the impurity. For times $t > 0$, we switch on the Kondo interaction and the state evolves nontrivially  as\be
\left|\Psi(t>0)\right\rangle=e^{-iH_{\rm eff}t}\left|\Psi_0\right\rangle.
\ee
By analogy with  the static problem \cite{Kondobook},  we expect   the infrared singularity associated with the Kondo effect to be cut off by the finite time after the impurity is coupled to the leads. Thus, the dynamics in the time regime $t\ll t_K$ must be  governed by the weak-coupling fixed point  and   $\lambda_K$ can be treated as a perturbative parameter. For times $t\sim t_K$, the dimensionless  Kondo  coupling must become of order 1,   implying that the perturbative expansion  breaks down. For $t\gg t_K$, the dynamics is controlled  by the strong-coupling fixed point. Besides the Kondo time scale, the  low-energy theory may involve another important time scale, $\Lambda_0^{-1}$, related to the microscopic details of the lattice model in  Eq.~(\ref{begin}). Since our quench is instantaneous,  the latter is  the shortest time scale  in the problem. The field theory results discussed  in the  following require $t\gg \Lambda_0^{-1} $, but we should still observe a crossover in the physical properties of the system between the intermediate-time regime $\Lambda_0^{-1}\ll t\ll  t_K $ and the long-time regime $t\gg t_K$.

\section{Charge transport across the impurity }
\label{sec4}

In this section we discuss the dynamics of the charge current after the impurity is coupled to the noninteracting chains with a small voltage bias.  Unless otherwise stated, we  assume  particle-hole symmetry  and set $W'=0$. 

\subsection{Weak coupling }

To study  time-dependent transport, we  consider different chemical potentials $\mu_{\ell}$ in the chains. The Hamiltonian for $t>0$ is modified by
\begin{equation}
    H \to H + \sum_{\ell=1,2}\mu_\ell N_\ell,
    \label{change}
\end{equation}
where $N_\ell$ is the total number operator for electrons in  chain $\ell$. The chemical potential term in Eq.~(\ref{change}) can be traded for a time-dependent vector potential using   the gauge transformation $\psi_{\ell} \to e^{-i\mu_{\ell}t}\psi_{\ell}$,   at  the price of introducing an explicit time dependence in $H$ \cite{Schiro2015}.

The current operator in the continuum limit  is 
\begin{eqnarray}
    \hat j(t)&=&e\frac{d}{dt}(N_1 - N_2)=ie[H_K,N_1-N_2],\nonumber\\
    &=&ie\pi v_F\lambda_K\left[ e^{ieV t}\mathbf S_0\cdot \psi^{\dagger}_2(0) \boldsymbol\sigma\psi^{\phantom\dagger}_1(0) - \text{h.c.}\right],
    \label{joper}
\end{eqnarray}
where $V\equiv(\mu_2-\mu_1)/e$ plays the role of a  bias voltage. In the following  we focus on the linear response regime and assume that $eV$ is the smallest energy scale in the problem, therefore all our time-dependent transport results are limited by the time $t\ll t_{V}\equiv|eV|^{-1}$. The current at time $t>0$ is given by  
 \begin{eqnarray}
      j(t)&=&\langle\Psi(t)|\hat j(t)|\Psi(t)\rangle,\nonumber\\
     &=&\langle  T\hat j_I(t)\:\mc U_{\text{Kel}} \rangle_0,
     \label{jresult}
\end{eqnarray}
where $\hat j_I(t) = e^{iH_0t}\hat j(t)e^{-iH_0t}$ is the current operator evolved in the interaction picture, $H_0$ is written in the continuum limit as in Eq. (\ref{Hwire}), and $\langle\,\rangle_0$ denotes the expectation value in  $  | \Psi_0\rangle$. Here $T$ is the time-ordering operator in Keldysh contour $\gamma$ and
\begin{equation}
    \mc U_{\text{Kel}} = \exp\left[-i\int_{\gamma}dt'H_{K,I}(t')\right]
    \label{U}
\end{equation} 
is the time evolution operator  in  $\gamma$ \cite{keldysh} involving the Kondo interaction $H_{K,I}(t)= e^{iH_0t}H_Ke^{-iH_0t}$ with $H_K$ given in Eq. (\ref{JK1}).

In the time regime $t\ll t_K$, the Kondo coupling can be treated perturbatively.  Expanding the exponential in Eq.~(\ref{U}) in powers of $\lambda_K$ and rewriting $\int_\gamma = \int_{\gamma^+} + \int_{\gamma^-}$, with $\gamma^-$ ($\gamma^+$) the time (anti-time)-ordered branch, we obtain a perturbative series for the current in Eq.~(\ref{jresult}).  The lowest-order terms in this series are 
\bea
j^{(2)}(t) &=& i\int_0^t dt'\langle [ H_{K,I}(t'),\hat j_I(t)] \rangle_0,
\label{O2}\\
j^{(3)}(t) &=&-\frac{1}{2}\int_0^t dt'dt''\Big[\langle \hat j_I(t)T H_{K,I}(t')H_{K,I}(t'') \rangle_0\nonumber\\
&&+ \text{h.c.}-2 \langle H_{K,I}(t')\hat j_I(t)H_{K,I}(t'') \rangle_0 \Big]. 
\label{O3}
\eea
Here $j^{(n)}(t)$ stands for the current at order $\lambda_K^n$, generated by expanding the time evolution operator to order $\lambda_K^{n-1}$.  Note that  the current operator     in Eq. (\ref{joper}) already contains one factor of $\lambda_K$. Since $[H_0,\mb S_0] = 0$, the    time dependence of the impurity spin correlators only comes from the time-ordering operator. 

Calculating  the correlator in Eq.~(\ref{O2}), we obtain  the    linear-in-$V$ contribution 
\begin{eqnarray}
j^{(2)}(t)&=&\frac{3ie^2V\lambda^2_K}{8}\int_0^{\Lambda_0t} du\,\frac{u}{[h(u)]^2} + \text{h.c.},
\label{OJ2}
\end{eqnarray}
where  $h(u) = 1+i u$. While the above integral can be computed analytically for arbitrary times, we are mostly interested in the regime  $\Lambda_0 t\gg 1$.  In this case, the leading term in the current behaves as \be
j^{(2)}(t) \approx \frac{3\pi^2j_0}{8}\lambda^{2}_K\left[1-\frac{4}{\pi \Lambda_0t}+\mc O(t^{-3})\right],\label{j2}\ee
 where $j_0=e^2V/\pi$ is the current for an ideal conductance.  This result is the same as that for the transport across a non-magnetic impurity \cite{Schiro2015}. As it stands, this result suggests that the current would approach a finite value corresponding to a small conductance of order $\lambda_K^2$. 

To   capture the Kondo effect,  we need to include the contribution to the current at order   $\lambda^3_K$.   Given the initial  state in Eq. (\ref{Psi0}) and   the current operator in Eq.~(\ref{joper}), the only   impurity correlator that contributes to the current  in Eq.~(\ref{O3})  is 
\begin{equation}
    \langle T S^a_0(t)S^b_0(t')S^c_0(t'') \rangle_0 =\epsilon^{abc} \frac{i}{8}\epsilon(t,t',t''),
    \label{s+s-sz}
\end{equation}
where $\epsilon^{abc}$ is the Levi-Civita symbol and $\epsilon(t,t',t'') = 1$ if $t>t'>t''$ and is completely antisymmetric under the exchange of $t,t',t''$. As for the conduction electrons, the   nonzero correlators involve, for instance,   
\begin{equation}
    \langle T\psi^{\dagger}_{1\downarrow}(t)\psi^{\phantom\dagger}_{2\uparrow}(t)\psi^{\dagger}_{1\uparrow}(t')\psi^{\phantom\dagger}_{1\downarrow}(t')\psi^{\dagger}_{2\uparrow}(t'')\psi^{\phantom\dagger}_{1\uparrow}(t'') \rangle_0,
    \label{co1}
\end{equation}
\begin{equation}
    \langle T\psi^{\dagger}_{1\downarrow}(t)\psi^{\phantom\dagger}_{2\uparrow}(t)\psi^{\dagger}_{2\uparrow}(t')\psi^{\phantom\dagger}_{1\downarrow}(t')\left[\rho_{1s}(t'') + \rho_{2s}(t'')\right]\rangle_0,
    \label{co2}
\end{equation}
where $\rho_{\ell s}=(\rho_{\ell\uparrow} - \rho_{\ell\downarrow})/\sqrt{2}$ is  defined in terms of   $\rho_{\ell \sigma}=\,:\psi^\dagger_{\ell\sigma}\psi^{\phantom\dagger}_{\ell\sigma}:$ and all the fields   act at $x=0$. After integrating    over $t''$ in Eq.~(\ref{O3}), we obtain   to linear order in $V$
\begin{eqnarray}
j^{(3)}(t) &=& \frac{ie^2 V\lambda^3_K}{4 }\int_0^{\Lambda_0t} du\frac{u\ln h(u)}{[h(u)]^2}\left\{ 1 -\frac{2[h(u)]^2}{ u( u -2i)} \right\}\nonumber\\
&&+ \text{h.c.}+\dots,
\label{OJ3}
\end{eqnarray}
where we drop terms that decay as  $ (\Lambda_0t)^{-1}$ or faster for $\Lambda_0t\gg1$. This approximation is equivalent to taking  the scaling limit of the Kondo model, $\Lambda_0\to \infty$, $J_K\to0$,  with $T_K$ fixed.  

Evaluating the integral in Eq. (\ref{OJ3}) and combining the result with the leading contribution from Eq. (\ref{j2}), we  obtain  
\begin{equation}
    j(t) \approx  \frac{3\pi^2j_0}{8}\lambda^{2}_K\Big[ 1 + 2\lambda_K\ln (\Lambda_0t) \Big] + \mc O(\lambda^4_K).
    \label{wfree}
\end{equation}
In this perturbative  regime,  the time dependence of the current can be cast in  the   form $j(t)=j[\lambda^{\rm eff}_K(t)]$, where  $j[\lambda_K]= \frac{3\pi^2j_0}{8}\lambda^{2}_K$ and
\begin{equation}
    \lambda^{\rm eff}_K(t) = \lambda_K + \lambda^{2}_K\ln (\Lambda_0 t) + \dots
    \label{Keff}
\end{equation}
is the effective  Kondo coupling at time scale $t$.  This result confirms our   expectation that the weak-coupling expansion should break down at long times, since the effective Kondo coupling diverges for $t\to\infty$. From Eq.~(\ref{Keff}), we  can define  the Kondo time by the condition $\lambda^{\rm eff}_K(t_K)\sim 1$ with bare coupling  $\lambda_K=\lambda_0\ll1$. This condition gives $t_K\sim \Lambda_0^{-1}e^{1/\lambda_0}\gg \Lambda_0^{-1}$, as expected from the relation  $t_K\sim 1/T_K$.  From Eq. (\ref{efflambda}), we then  have $\lambda^{\rm eff}_K(t)\approx [\ln(t_K/t)]^{-1}$  for $\Lambda_0^{-1}\ll t\ll t_K$,   and we obtain \be
 j(t) \approx  \frac{3\pi^2}{8 } \frac{j_0}{\ln^2(t_K/t)}.
\ee 
This result was obtained in Ref. \cite{Pletyukhov2010} based on poor man scaling's arguments. Here we have explicitly verified the scaling of the effective Kondo coupling by computing the third-order contribution to the time-dependent current.

The scattering potential term in Eq.~(\ref{JK1}) contributes to   the current at order $(W')^2$  \cite{Schiro2015}. Since this term does not renormalize, the result   in Eq.~(\ref{wfree}) is simply modified by a constant term.  Similarly to the low-energy limit of the Kondo problem in the static case, the long-time limit of our dynamical problem is governed by the flow of $\lambda_K$ to strong coupling.  We will address this limit in the next subsection.

\subsection{Strong coupling\label{sec:sc}}

In the limit $t\gg t_K$, the effective Kondo coupling  diverges. The impurity spin is completely screened and removed from the low-energy theory. In the particle-hole symmetric case,  $W'=0$, the scattering phase shift associated with the Kondo effect in the symmetric channel  \cite{Pustilnik, Affleck2010} modifies the boundary conditions at the origin to \be
\psi_{L,2,\sigma}(0)=\psi_{R,1,\sigma}(0),\qquad \psi_{R,2,\sigma}(0)=\psi_{L,1,\sigma}(0).
\ee
Thus, the  fixed-point Hamiltonian describes a single wire with perfect transmission at the origin: 
 \be
H_{\text{sc}}= v_F\int dx\, [\psi_{R}^\dagger(-i\partial_x)\psi_R+ \psi_{L}^\dagger(i\partial_x)\psi_L]. 
\ee 
As a consequence, for sufficiently long times  the current must approach the ideal conductance limit,  $j(t\gg t_K) \to j_0$.   

\begin{figure}
    \centering
    \includegraphics[width=0.9\columnwidth]{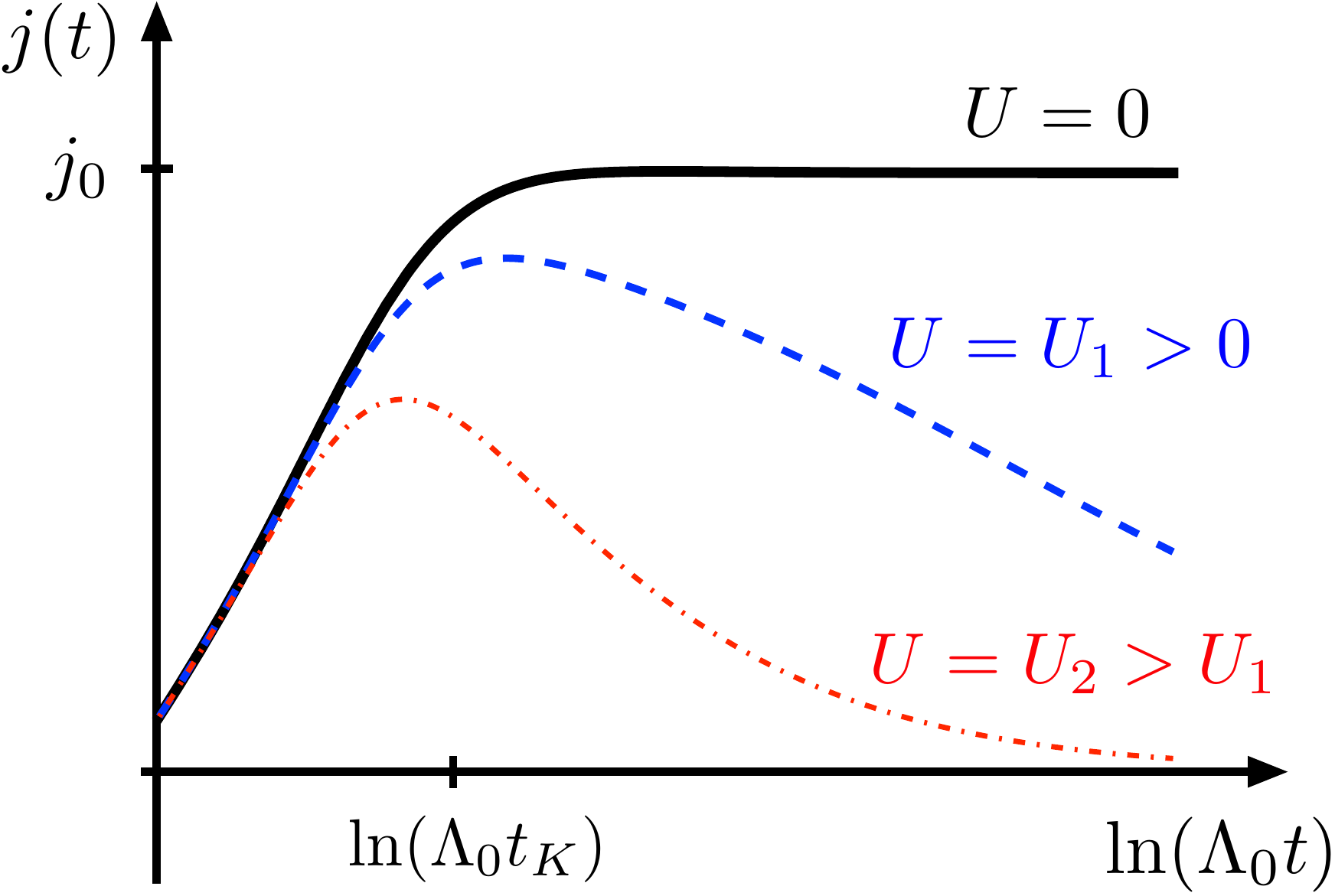}
    \caption{Schematic time dependence of the current across the impurity spin after the quantum quench. Here $j_0=G_0V$ is the current for ideal conductance $G_0=e^2/\pi$. The solid line represents the case of noninteracting electrons in the chains. In the  time interval  $\Lambda_0^{-1}\ll t\ll t_K$, described by the weak-coupling fixed point of the Kondo model, the current increases logarithmically with time, see Eq.  \eqref{wfree}. The long-time regime $t\gg t_K$ is described by the strong-coupling fixed point, see Eq. (\ref{scfree}). The dashed and dot-dashed lines represent the behavior for two different values of the repulsive electron-electron interaction in the leads, in which case the current vanishes as $\sim t^{1-K_c^{-1}}$ at long times [see Eq. (\ref{winterac}) and Sec. \ref{SCint}].}
    \label{figure2}
\end{figure}

 To analyze the long-time behavior,   we can perturb the strong-coupling fixed point by its leading irrelevant operator.  Since the impurity is screened, its spin operator does not appear in the low-energy effective Hamiltonian. According to local Fermi liquid theory \cite{Nozieres1974, Affleck2010}, the leading irrelevant operator  that respects SU(2) symmetry is 
\begin{equation}
    \delta H = g :\mathbf J^2(0):,
    \label{J^2}
\end{equation}
where $\mathbf J(0) = \psi^{\dagger}(0)\frac{\bm{\sigma}}{2}\psi(0)$ is the spin density at the origin and $ g \sim 1/T_K \sim t_K$ is the   coupling constant. The operator   in Eq.~(\ref{J^2}) has scaling dimension equal to $2$; like  any boundary operators with dimension greater than 1, $g$ is irrelevant in the RG sense. Considering the application of a small voltage $V$ around the origin \cite{ Giamarchi} and calculating the current at long times by perturbation theory  to order $g^2$, we obtain
\begin{equation}
    j(t) \approx j_0\left[ 1 -\left( \frac{t_K}{t} \right)^2\right].
    \label{scfree}
\end{equation}
Thus, in the long-time regime the Kondo time $t_K$ can be extracted from the coefficient of the $1/t^2$ term, which is absent in the intermediate-time regime, see Eq. (\ref{j2}). We recall that we always consider   the time range $t\ll t_{V} =|eV|^{-1}$. At finite bias,  the current would be  influenced by oscillating terms with frequency $eV$ (see e.g. Ref. \cite{Pletyukhov2010}).

The   results   in Eqs.~(\ref{wfree}) and (\ref{scfree}) are illustrated in Fig. \ref{figure2}. They are analogous to   the behavior of  static properties of   Kondo systems at finite temperature, such as the impurity susceptibility $\chi_{\rm imp}(T)$ \cite{Kondobook},  with the correspondence $T \leftrightarrow t^{-1}$. This correspondence  reveals that, in  the post-quench dynamics,    time essentially acts  as the inverse of an energy scale and  the dynamics is effectively controlled  by the RG flow  from weak to strong coupling.     
At intermediate times, $t\sim t_K$, the problem becomes non-perturbative and we are not able to derive analytical expressions. However, since the Kondo effect is characterized by a crossover with a single  emergent energy scale $T_K$, in general we expect   a smooth function $j\left(t/t_K\right)$   connecting the asymptotic behavior in Eqs.~(\ref{wfree}) and (\ref{scfree}).

For a system without particle-hole symmetry, we need to take into account  an additional perturbation corresponding to a  local potential barrier: 
\begin{equation}
    H_{\text{sc}} \to H_{\text{sc}} + \mc W:[\psi_R^{\dagger}(0)+\psi_L^{\dagger}(0)][\psi^{\phantom\dagger}_R(0)+\psi^{\phantom\dagger}_L(0)]:,
    \label{scfp}
\end{equation}
where  $\mc W$ is the strength of the scattering potential.  For noninteracting electrons, this term is a marginal perturbation. As a result,   the  constant value of the current at long times is reduced  by a correction   of order $\mc W^2$, but the time dependence is qualitatively the same as for $ \mc W=0$.

\section{Luttinger liquid effects}
\label{sec5}

In this section we discuss how electron-electron interaction in the chains affect the Kondo effect in the post-quench dynamics.  In this case, our model describes two semi-infinite Hubbard chains prepared in the ground state of $H_0+H_U$ and suddenly coupled to a polarized impurity spin.  

\subsection{Low-energy Hamiltonian}

The low-energy excitations in the chains are described by the Luttinger model  \cite{ Giamarchi,gogolin}. Starting   from $t'_1=0$ (see Fig. \ref{system}), we   take the continuum limit and write the interaction in terms of the chiral fermionic fields. The low-energy effective Hamiltonian for each    chain reads
\begin{eqnarray}
H_{\ell}^{\rm int}&=&\int dx\,\Big\{v_F\psi^\dagger_{\ell}(-i\partial_x)\psi^{\phantom\dagger}_{\ell} \nonumber\\
&&+U\sum_{\sigma}:\psi^{\dagger}_{\ell,\sigma}(x)\psi^{\dagger}_{\ell,-\sigma}(-x)\psi_{\ell,\sigma}(-x)\psi_{\ell,-\sigma}(x): \nonumber\\
&& + U\sum_{\sigma} \rho_{\ell,\sigma}(x)[\rho_{\ell,-\sigma}(x) +\rho_{\ell,-\sigma}(-x)]\Big\}, 
\label{HU}
\end{eqnarray}
where  we used  the constraint   in Eq.~(\ref{constr}). Here we have omitted the Umklapp term, which   oscillates in space  for $k_F\neq \pi/2$  \cite{Giamarchi}.   At half-filling, $k_F=\pi/2$,  the Umklapp term becomes a relevant perturbation that  drives the system to a Mott-insulating phase for arbitrarily small $U>0$. For this reason, in the interacting case we   stay away from half-filling, which entails breaking particle-hole symmetry. 

We can diagonalize the interacting   Hamiltonian in Eq.~(\ref{HU})   using Abelian bosonization   \cite{Giamarchi, gogolin}. For   open boundary conditions, the fermionic field operator assumes the form \cite{Fabrizio1995}
\begin{equation}
\psi_{\ell}(x)\sim\frac{1}{\sqrt{2\pi\alpha}}\left(\begin{array}{c}F_{\ell\uparrow}e^{-i\sqrt{\frac{\pi}{2}}[\phi_{\ell,c}(x)+\phi_{\ell,s}(x)]}\\
F_{\ell\downarrow}e^{-i\sqrt{\frac{\pi}{2}}[\phi_{\ell,c}(x)-\phi_{\ell,s}(x)]} \end{array} \right),  \label{boso}
\end{equation}
where  $\alpha\sim v_F/\Lambda_0$ is a short-distance cutoff, $F_{\ell\sigma}$ are   Klein factors  that ensure the anticommutation relations between electrons with opposite spin, and $\phi_{\ell,\lambda}(x)$ with $\lambda=c,s$  are   chiral bosonic fields associated with charge and spin collective modes,  which obey 
\begin{equation}
[\phi_{\ell,\lambda}(x),\phi_{\ell',\lambda'}(y)]=i\delta_{\ell\ell'}\delta_{\lambda\lambda'}\sgn(x-y).
\end{equation}
In terms of bosonic annihilation operators $\eta_{\ell,\lambda, q}$ with momentum $q>0$, the fields $\phi_{\ell,\lambda}(x)$ are given by 
\begin{equation}
\phi_{\ell,\lambda}(x)= \sum_{q>0}\frac{e^{-\frac{\alpha }{2}q}}{\sqrt{qL}}[ z_{\lambda q}(x)\eta^{\phantom\dagger}_{\ell,\lambda, q} + z^{*}_{\lambda q}(x)\eta^{\dagger}_{\ell,\lambda, q}],    \label{modeexp}
\end{equation}
where $L$ is the length of the open chain,  $z_{\lambda q}(x) = (1/\sqrt{K_{\lambda}}) \cos(qx) + i\sqrt{K_{\lambda}}\sin(qx)$, and $K_c$ and $K_s$ are the Luttinger parameters in the charge and spin sectors, respectively. For the Hubbard model with $U\geq0$,  we have $1/2 < K_c \leq 1$, with $K_c=1$ corresponding to the free-fermion point. In the spin sector, the SU(2)  spin-rotation   symmetry fixes $K_s=1$  \cite{Giamarchi, gogolin}.

The bosonized   Hamiltonian for an interacting chain in Eq.~(\ref{HU}) can be written as
\begin{equation}
H^{\rm int}_{\ell}= H_{\ell}^{\rm LL} + H_\ell^{\text{bs}},
\label{LL}
\end{equation}
where the first term describes a Luttinger liquid with open boundary conditions:
\begin{equation}
    H_{\ell}^{\rm LL} = \sum_{\lambda=c,s}\sum_{q>0}v_{\lambda}q\,\eta^{\dagger}_{\ell,\lambda, q}\eta^{\phantom\dagger}_{\ell,\lambda, q}.
    \label{Lut}
\end{equation}
Here $v_{c}$ and $v_s$ are the velocities of the charge and spin bosonic modes, respectively. While  bosonization yields perturbative expressions for the velocities and Luttinger parameters for small $U$, the Luttinger liquid Hamiltonian holds in general  as  the low-energy fixed point for any metallic (gapless) system in one dimension \cite{Haldane2}. The second term in Eq. (\ref{LL}) corresponds to the backscattering operator in the  second line of Eq.~(\ref{HU}). For the SU(2)-symmetric model with $U>0$, this term is known to be marginally irrelevant \cite{Giamarchi, gogolin}.  In the following we neglect the effects of the backscattering term and approximate  $H^{\rm int}_{\ell}\approx H^{\rm LL}_{\ell}$.

We can now couple the quantum dot in the Kondo regime to the Luttinger liquid  leads \cite{Lee1992,Furusaki1994,Fabrizio1995,Frojdh1995}. We rewrite the  Kondo interaction as  \bea
H_K^{\rm int}&=&\pi v_F\lambda_K\mathbf S_0 \cdot\sum_{\ell}:\psi^{\dagger}_{\ell}(0)\frac{\boldsymbol\sigma}{2}\psi^{\phantom\dagger}_{\ell}(0):\nonumber\\
&&+\pi v_F\Gamma_K\mathbf S_0 \cdot\left[:\psi^{\dagger}_{1}(0)\frac{\boldsymbol\sigma}{2}\psi^{\phantom\dagger}_{2}(0)+\text{h.c.}\right]: . \label{JKint}
\eea
In addition,  we now distinguish between the dimensionless Kondo coupling within the same wire,  $\lambda_K$,  and the coupling $\Gamma_K$ associated with tunneling across the impurity because  these operators acquire different scaling dimensions in the interacting case.  The bosonization of the Kondo interaction can be obtained from Eq. (\ref{boso}). Close to the weak-coupling fixed point, the couplings  $\lambda_K,\Gamma_K \ll 1$ obey the RG equations \cite{Fabrizio1995}  
\bea
    \frac{d}{dl} \lambda_K&=& \frac{1}{2}(\lambda^2_K +\Gamma^2_K),
    \label{lambd}\\
    \frac{d}{dl} \Gamma_K&=& \frac12(1-K^{-1}_c)\Gamma_K + \lambda_K\Gamma_K.
    \label{Gama}
\eea
For a repulsive interaction, $K_c<1$, the coupling $\Gamma_K$ initially decreases under the RG flow. However, since $\lambda_K>0$ always increases, the second term in Eq.~(\ref{Gama})  ultimately drives $\Gamma_K$ to strong coupling as well.

The perturbative RG equations indicate that the low-energy limit of the Kondo model with interacting chains  is still described by a strong-coupling fixed point  where the impurity spin is screened by the conduction electrons. However, away from half-filling the scattering potential term in Eq. (\ref{scfp}) is allowed by symmetry as a perturbation to the fixed point with ideal conductance. Even if we assume a small $\mc W$, the backscattering part of  this term flows to strong coupling as $d\mc W/dl=(1-K_c)\mc W/2$. As a consequence, the effective height of the potential barrier diverges. At low energies, we recover two decoupled Luttinger liquids with open boundary conditions, with one electron in the symmetric channel  having been removed to form a singlet with the impurity spin. This picture for the   Kondo effect in a Luttinger liquid suggests that the time dependence of the current in our quench protocol can be strongly affected by interactions, as we shall discuss in the following.

\subsection{Weak coupling} 

We now turn to the post-quench dynamics for   interacting electrons. As discussed in Sec. \ref{sec2}, we assume that  the system has been prepared in the ground state of the Hamiltonian for decoupled chains. Within the low-energy theory, the ground state $\left|\psi_0\right\rangle_\ell$ of  the Hubbard chains corresponds to the vacuum of charge and spin bosons, $\eta_{\ell,\lambda,q}\left|\psi_0\right\rangle_\ell=0$ for $\lambda=c,s$ and  all $q>0$.

Once again, we start by calculating the time-dependent current using   perturbation theory in the Kondo coupling.  The  first terms in the series expansion of $j(t)$ are still given by Eq. (\ref{O2}) and (\ref{O3}), but  we now use the bosonized expressions for the current operator and the Kondo interaction  given  in Appendix \ref{ap1}. The bias voltage is implemented as a time-dependent shift of the charge bosons, $\phi_{\ell,c}\to \phi_{\ell,c}+\sqrt{\frac2\pi}\mu_\ell t$. The correlators   involve exponentials of the bosonic fields, see Eq.~(\ref{boso}). At order  $\lambda^2_K$, we need   two-point functions of the form \cite{Giamarchi,gogolin}
\begin{equation}
\langle e^{ia\phi_{\ell,\lambda}(t)}e^{-ib\phi_{\ell',\lambda'}(t')} \rangle_0=\frac{\delta_{ab}\delta_{\ell\ell'}\delta_{\lambda\lambda'}}{[1+i\Lambda_0(t-t')]^{a^2/\pi K_{\lambda}}},\label{vertex}
\end{equation}
where $a \in \mathbbm R$. At order $\lambda^3_K$,   correlators such as the one  in Eq. (\ref{co1}) can be calculated using Eq. (\ref{vertex}). On the other hand, the bosonized form of the  correlator   in Eq.~(\ref{co2})  is proportional to  
\begin{equation}
    \langle T e^{-i\sqrt{\pi}[\phi^{+}_s(t)-\phi^-_c(t)]}e^{i\sqrt{\pi}[\phi^{+}_{s}(t')-\phi^-_c(t')]}\partial_x\phi^{+}_s(t'') \rangle_0,
    \label{C2}
\end{equation}
where we define the symmetric and antisymmetric combinations \be
\phi^{\pm}_{\lambda}(x)=\frac{\phi_{1,\lambda}(x)\pm \phi_{2,\lambda}(x)}{\sqrt{2}},\label{phipm}\ee
which obey $[\phi^{\sigma}_{\lambda}(x),\phi^{\sigma'}_{\lambda'}(y)]=i\delta_{\sigma\sigma'}\delta_{\lambda\lambda'}\sgn(x-y)$. To calculate the spin part of this correlator, we need to contract $\partial_x\phi^{+}_s$ with the exponentials  using Wick's theorem as explained in Appendix  \ref{ap2}. We have checked that our result agrees with  the  correlator obtained in fermionic language in the noninteracting limit $K_c\to1$.

After calculating the correlators, we find that the leading contributions to the  current in the interacting case  are obtained by replacing  $[h(u)]^2\to [h(u)]^{1+K_c^{-1}}$  in Eqs.~(\ref{OJ2}) and (\ref{OJ3}).  The Luttinger parameter $K_c$ also appears in the   subleading contributions omitted in Eq. (\ref{OJ3}).  For the time regime  $\Lambda_0 t\gg 1$, we can calculate the integrals analytically to obtain
\begin{eqnarray}
    j(t)&=&\frac{3\pi j_0}{4}\frac{\sin(\pi\nu/2)\Gamma^{2}_K}{\nu (\Lambda_0t)^{\nu}} \Big[ 1 + 2\lambda_K\ln (\tilde \Lambda_0t)  \Big] \nonumber\\
    && + \mc O(\Gamma^4_K),
    \label{winterac}
\end{eqnarray}
where $\nu=K_c^{-1}-1$, $\tilde\Lambda_0 =\Lambda_0 \exp\left[\frac{1}{\nu} -\frac{\pi}{2}\cot(\frac{\pi\nu}{2})\right]$ and we assumed $\alpha\Lambda_0 \approx v_F$. It is easy to verify  that  Eq.~(\ref{winterac}) reduces to  Eq.~(\ref{wfree})   if we   take  $\Gamma_K=\lambda_K$ and $K_c\to1$. In the above expression we can recognize the competing effects in the intertwined RG equations  (\ref{lambd}) and (\ref{Gama}).  At second order, we find a power-law decrease in the effective tunneling amplitude $\Gamma_K$, but  at third order there is a logarithmic enhancement due to the coupling between $\Gamma_K$ and   $\lambda_K$.

We can rewrite Eq.~(\ref{winterac}) in terms of a renormalized coupling as
\begin{equation}
  j(t) = \frac{3\pi j_0}{4}\frac{\sin(\pi\nu/2)}{\nu }[\Gamma^{\rm eff}_K(t)]^{2},
\end{equation}
  where
\begin{equation}
    \Gamma^{\rm eff}_K(t) \approx  \frac{1}{(\Lambda_0 t)^{  \nu/2   }}\frac{\Gamma_K}{1-\lambda_K\ln (\tilde \Lambda_0 t)}.
    \label{GamRenor}
\end{equation}
Besides two independent Kondo  couplings, $\Gamma_K$ and $\lambda_K$, the expression in Eq. (\ref{GamRenor}) involves the high-energy cutoff in the power-law decaying factor.  The interactions in the wires destroy the universal scaling of the Kondo problem in the sense that  it is no longer possible to write the current, or any other physical quantity, as a function of $t/t_K$ only. Nevertheless, it is instructive to analyze the weakly interacting limit $U\ll t_1$, where we have $K_c\approx 1- U/(\pi v_F)$  \cite{Giamarchi}. In this case, $\lambda_K$ and $\Gamma_K$ start off  approximately equal and flow together to strong coupling.  Perturbation theory breaks down when $\lambda_K\ln (\tilde \Lambda_0 t)\sim 1$, which gives an estimate for the time scale\begin{equation}
    \tilde t_K \sim t_K\left(1- \frac{\pi U}{12v_F}\right),
    \label{KT2}
\end{equation}
where $t_K=\Lambda_0^{-1}e^{1/\lambda_K}$ is the Kondo time in the noninteracting case and we have expanded the renormalized cutoff to first order in $U$. 
In principle, at higher orders  the dimensionless  Kondo coupling $\lambda_K$ may pick up a $U$-dependence as well,   but in the weakly interacting limit  this dependence can be ignored   in Eq. (\ref{KT2}). According to Eq.~(\ref{KT2}), the time $\tilde t_K$ decreases with the interaction in the wires; in other words, repulsive  interactions in the wires speed up the formation of Kondo singlet state. The same conclusion was reached numerically in Ref. \cite{Helena}.

Since particle-hole symmetry is broken away from half-filling, we also need to consider the  perturbation in Eq.~(\ref{HW}).   In bosonized form, the intrawire part of this potential scattering term is proportional to  $\partial_x\phi^+_c(0)$ and produces a constant contribution to the current. On the other hand,  the tunneling term is proportional to  $\cos[\sqrt{\pi}\phi^-_c(0) +eVt]\cos[\sqrt{\pi}\phi^-_s(0)]$, yielding a   contribution that decays with time  as $1/t^\nu$, as expected from the scaling dimension of this operator.     
 
\subsection{Strong coupling}
\label{SCint}
We now turn to  the long-time regime $t\gg \tilde t_K$. As discussed in Sec. \ref{sec:sc},  this regime is governed by perturbations to the strong-coupling fixed point. The main difference from the noninteracting case is that  the potential scattering  term in Eq.~(\ref{scfp}) now has scaling dimension  $(1+K_c)/2 $ and becomes  relevant  for repulsive interactions. In this case, the stable low-energy fixed point of the Kondo model  consists of two decoupled semi-infinite Luttinger liquids \cite{Fabrizio1995}. Since this fixed point has vanishing conductance, we expect $j(t\to \infty)=0$. The leading perturbation   is the irrelevant tunneling between the wires, equivalent to  the interwire term in Eq.~(\ref{HW}). Thus,  we conclude that, after the impurity spin has been effectively screened, the current goes to zero as $j(t) \sim1/t^{\nu}$. This behavior is represented by the dashed and  dot-dashed lines in Fig. \ref{figure2}.

\section{Impurity magnetization}
\label{magneSec}

In this section we consider  the time evolution of the impurity magnetization after the quantum quench. The purpose  is to connect with the results of Ref. \cite{Helena}, where second-order perturbation theory in the Kondo coupling was used to analyze numerical data from   time-dependent DMRG.  Here we extend this calculation to third order and contrast   the behavior of impurity magnetization with that of the charge current discussed in Sec. \ref{sec4}.

We now consider the simpler setup in which  the impurity is   coupled at the edge of a single  chain, say   $\ell =2$ in Fig. \ref{system}.  In this case the Kondo interaction   is 
\begin{eqnarray}
H'_K&=&  J_K\mathbf S_0\cdot c^\dagger_{1}\frac{\boldsymbol\sigma}{2}c^{\phantom\dagger}_{1},
\end{eqnarray}
where  $J_K$ is still given by  Eq.~(\ref{JK}). To obtain the low-energy effective Hamiltonian, we   proceed  as in Sec.~\ref{sec5}. Using the bosonization mapping and neglecting  the irrelevant backscattering term in the interacting case, we obtain  the effective Hamiltonian 
\begin{eqnarray}
   H'&=&H^{\rm LL} + \pi v_s \lambda_{K}\left[{S}^{+}_{0}\frac{e^{-i\sqrt{2\pi}\phi_{s}(0)}}{2\pi\alpha}+ {S}^{-}_{0}\frac{e^{i\sqrt{2\pi}\phi_{s}(0)}}{2\pi\alpha}\nonumber\right.\\
&&\left. -  {S}^{z}_{0}\frac{1}{\sqrt{2\pi}}\partial_x\phi_{s}(0)\right],\label{HK2}
\end{eqnarray}
where we drop the index $\ell$ in the fields for a single chain. Remarkably, in the above Hamiltonian the Kondo interaction  only involves the spin boson  $\phi_s$. This  is a result of spin-charge separation  in the geometry with an impurity coupled to the  edge of a single  wire \cite{Pereira2008}. Thus, unlike the case of an impurity embedded between two wires discussed in the previous sections, in this case  the charge sector remains free and, as consequence, the Luttinger parameter $K_c$ does not appear in correlators for the electron spin density.

The impurity magnetization  is given by \be
m_0(t)=\langle \Psi (t)|S^z_0|\Psi(t)\rangle.\ee
We can apply  perturbation theory  in $\lambda_K$ to obtain an expression for $m_0(t)$ valid in the regime $t\ll t_K$. Alternatively, we can employ a  description similar to  that used in the calculation of the charge current  and consider the quantity [cf. Eq. (\ref{joper})]
\begin{equation}
\hat j^z_s = \frac{d}{dt}S^z= \pi v_s \lambda_K :\psi^{\dagger}(0)(\mathbf S_{0}\times\bm{\sigma})_z\psi(0):,
\label{spincur}
\end{equation}
where $S^z$ is the $z$ component of  the total spin operator for electrons in the wire. The operator in Eq.~(\ref{spincur}) represents the spin current that flows to the wire after the quench. In terms of the bosonic fields, the spin current reads  
\begin{equation}
   \hat  j^z_s = i\frac{v_s\lambda_K}{2\alpha}\left[ S^+_0e^{-i\sqrt{2\pi}\phi_s(0)} - S^-_0e^{i\sqrt{2\pi}\phi_s(0)} \right].
\end{equation} 
 Since the entire system conserves spin, the impurity magnetization is related to the spin current for  $t>0$ by
\begin{equation}
     j^z_s(t)   + \frac{d}{dt}m_0(t) = 0,
    \label{continu}
\end{equation}
where $   j^z_s(t) =\langle \Psi(t)| \hat j^z_s|\Psi(t)\rangle$. This continuity equation allows us to obtain $m_0(t)$ by integrating $j^z_s(t)$ with the initial condition $m_0(0)=1/2$.

We can now calculate $ j^z_s(t)$ by perturbation theory  in $\lambda_K$. The  expansion is analogous  to Eqs.~(\ref{O2}) and (\ref{O3}).  At order $\lambda^2_K$, we find  that $ j^z_s(t)$ decays as $1/t$ for $\Lambda_0 t\gg 1$. Integrating this leading contribution in time according to Eq.  (\ref{continu}), we recover the  logarithmic relaxation of the impurity magnetization observed in Ref.  \cite{Helena}. Going further and computing the third-order contribution, we obtain  \begin{equation}
      j^z_s (t) = \frac{\lambda^{2}_K}{2t}\left[  1 + 2\lambda_K\ln (\Lambda_0 t) \right] + \mc O(\lambda^4_K),
    \label{spincur1}
\end{equation}
where we drop a subleading contribution that decays as $1/t^2$ in the regime $\Lambda_0t\gg 1$. Thus,  the spin current can be cast in the form $ j^z_s (t) = [\lambda^{\rm eff}_K(t)]^2/(2t)$ with  $\lambda^{\rm eff}_K(t)$ given  in  Eq.~(\ref{Keff}). On the other hand, using  Eq.~(\ref{continu}), we find that the impurity magnetization is given by 
\begin{equation}
    m_0(t) = \frac{1}{2} - \frac{\lambda_K^2}{2}\ln (\Lambda_0 t) \left[1 +  \lambda_K\ln (\Lambda_0 t) \right] + \mc O(\lambda^4_K).
    \label{magne}
\end{equation}
We note that $m_0(t)$ is not   a function of the effective Kondo coupling and, consequently,   not a function of $t/t_K$. In hindsight, this conclusion could have been anticipated by noting that   the second-order term in Eq. (\ref{magne})  contains an explicit logarithmic dependence on the high-energy cutoff $\Lambda_0$ which is not related to the Kondo effect. We can further interpret this result using a simple scaling argument. The continuity equation  implies that the current associated with any dimensionless conserved charge has dimensions of energy (or inverse time). For the charge current obtained in Eq. (\ref{wfree}), the factor with dimensions of energy stems from the voltage bias $eV$ that drives the current. In this case, the leading  time dependence in the scaling limit enters in the renormalization of the dimensionless Kondo coupling. By contrast, the spin current in Eq. (\ref{spincur1}) is injected into the chain due to the relaxation of the impurity spin  in the absence of perturbations such as  a spin chemical potential bias. As a result of the scale invariance of the unperturbed system, the dimension contained in $j_s^z(t)$ appears as an explicit time dependence in the  factor of $1/t$. This factor diverges for $t\to 0$, which implies that the integral of the spin current over time   must include the  short-time cutoff   $\Lambda_0^{-1}$. This effect accounts for  the additional cutoff dependence in  the result of Eq. (\ref{magne}).

Finally, we note that Eq.~(\ref{magne}) also describes  the impurity magnetization in the embedded geometry of Fig. \ref{system} for $U=0$ since in this case we can rewritten the Kondo Hamiltonian only in terms of a single noninteracting channel $\psi_{+}$, where $\psi_{\pm}=(\psi_1 \pm \psi_2)/\sqrt{2}$ \cite{Simon2001}. In the interacting case, the expression for $m_0(t)$ in the embedded geometry contains a contribution that decays as  $1/t^\nu$ in order $\Gamma^2_K$ \cite{Helena}.

\section{CONCLUSION}
\label{concluSec}

We studied a quantum quench in which a magnetic impurity is suddenly coupled to  the boundary of open chains. We focused on the real-time post-quench dynamics of two observables, the tunneling current across the impurity and the impurity magnetization. For noninteracting electrons, we found that the time-dependent current is a scaling function of the ratio $t/t_K$.  The  regimes $t\ll t_K$ and $t\gg t_K$ are governed by the weak- and strong-coupling fixed points of the Kondo model, respectively. For interacting chains, the current at intermediate times exhibits  a power-law dependence characteristic of Luttinger liquids physics and a logarithmic enhancement associated with the Kondo effect. Analyzing the weakly interacting case, we found that   repulsive interactions  decrease the Kondo time scale and favor  the formation of the Kondo cloud,  in accordance with  previous numerical results~\cite{Helena}. Concerning the impurity magnetization, we concluded that the latter  is not a   function of the renormalized Kondo coupling  in the   regime  $t\ll t_K$ because   the perturbative result contains  an   explicit dependence on the short-time cutoff $\Lambda_0^{-1}$ which is not related to the Kondo effect. 

Our results emphasize the importance of identifying suitable physical properties when searching for Kondo physics in the real-time evolution. The scaling limit  requires $t_K\gg \Lambda_0^{-1}$,   making it challenging to observe the crossover from short to long times in currently available numerical simulations. Some advantage may be gained by considering   time-dependent spin transport in the spin chain version of the Kondo model \cite{Eggert1992,Laflorencie2008,Giuliano2017}.  In this case,  the charge degree of freedom is gapped out, and the model can be realized in the Mott-insulating phase of bosonic atoms in deep optical lattices. An important difference is that  a magnetic  impurity embedded between two Heisenberg  spin chains gives rise to the two-channel Kondo effect, with a non-Fermi-liquid fixed point that may be manifested in the long-time post-quench dynamics. We leave this problem as a possible direction for future work.

\acknowledgements
This work is supported by FAPEMIG, CNPq (in particular through INCT- IQ 465469/2014-0), and CAPES (in particular through program CAPES-COFECUB-0899/2018). Research at IIP-UFRN is supported by Brazilian ministries MEC and MCTI. 

\appendix
 

\section{Bosonization formulas} \label{ap1}

In this appendix we write down the bosonized expressions for some important quantities in the interacting case.

After the gauge transformation $\psi_{\ell} \to e^{-i\mu_{\ell}t}\psi_{\ell}$ the Kondo Hamiltonian in Eq.~(\ref{JKint}) and the current operator defined in Eq.~(\ref{joper}) are given by

\begin{equation}
    H_K^{\rm int}(t) = \pi v_F\lambda_K[S^+_0F(t) + S^-_0F^{\dagger}(t) + S^z_0G(t)],
\end{equation}
\begin{equation}
    \hat j(t) = ie\pi v_F\Gamma_K[S^+_0 f(t) - S^-_0f^{\dagger}(t) + S^z_0g(t)],
\end{equation}
where
\begin{equation}
    f(t) = \frac{-i}{\pi\alpha}e^{-i\sqrt{\pi}\phi^{+}_s(0)}\sin[eVt + \sqrt{\pi}\phi^{-}_c(0)],
\end{equation}
\begin{equation}
    g(t) = \frac{-2i}{\pi\alpha}\sin[\sqrt{\pi}\phi^{-}_s(0)]\cos[eVt + \sqrt{\pi}\phi^{-}_c(0)],
\end{equation}
\begin{eqnarray}
F(t)&=&\frac{1}{2\pi \alpha}e^{-i\sqrt{\pi}\phi^+_s(0)}\Big\{\cos[\sqrt{\pi}\phi^-_s(0)]\nonumber\\
&&+\mc C\cos[\sqrt{\pi}\phi^{-}_{c}(0)+eVt]\Big\},
\end{eqnarray}
\begin{eqnarray}
  G(t)&=&-\frac{1}{\sqrt{4\pi}}\partial_x\phi^{+}_s(0) \nonumber \\
  && - \frac{\mc C}{\pi \alpha}\sin[\sqrt{\pi}\phi^{-}_c(0)+eVt]\sin[ \sqrt{\pi}\phi^{-}_s(0)],
\end{eqnarray}
where $\phi^{\pm}_{\lambda}(x)$ are defined in Eq. (\ref{phipm}) and $\lambda_K\mc C \equiv \Gamma_K$. For a noninteracting system, we have $\lambda_K=\Gamma_K$, consequently $\mc C=1$. In the above expressions we do not explicitly write the Klein factors, but they must be taken into account when calculating the correlators.

\section{Three-point function} \label{ap2}

In this appendix, we present some details on the calculation of  the third-order correlator in Eq.~(\ref{C2}). The charge part of the correlator reduces to a  two-point function as given in Eq. (\ref{vertex}). The spin part   is equivalent to the bosonization  of the fermion operators in
\bea
    C&\equiv&\langle T\psi^{\dagger}_{\downarrow}(t)\psi_{\uparrow}(t)\psi^{\dagger}_{\uparrow}(t')\psi_{\downarrow}(t')\nonumber\\
    &&\times  [: \psi^{\dagger}_{\uparrow}(t'')\psi_{\uparrow}(t'') - \psi^{\dagger}_{\downarrow}(t'')\psi_{\downarrow}(t'') : ]\rangle.
    \label{I}  
\eea
For a noninteracting system, we can calculate this correlator by applying   Wick's theorem. Using the Green's function $\langle T\psi^{\dagger}_{\sigma}(t)\psi_{\sigma}(t') \rangle=\frac{-i}{2\pi v_F}\frac{1}{t-t'}$,   we obtain the result for $t>t'>t''$:
\begin{equation}
    C=-\frac{2i}{(2\pi v_F)^3}\frac{1}{(t-t')(t-t'')(t'-t'')}.
    \label{B1}
\end{equation}

We can now calculate the same correlator using the bosonization dictionary. For a system with open boundary conditions, we use   $\psi_{\sigma}(t) = \frac{F_{\sigma}}{\sqrt{2\pi\alpha}}e^{-i\sqrt{\pi}\phi_{\sigma}(t)}$ and $:\psi^{\dagger}_{\sigma}(t)\psi_{\sigma}(t): = -\frac{1}{\sqrt{4\pi}}\partial_x\phi_{\sigma}(t)$ for a  chiral boson that obeys $[\phi_\sigma(x),\phi_{\sigma'}(x')]=i\pi \delta_{\sigma\sigma'}\text{sgn}(x-x')$.  The correlator in Eq. 
(\ref{I}) becomes 
\begin{equation}
    C=-\frac{\langle Te^{-i\sqrt{2\pi}\phi_s(t)}e^{i\sqrt{2\pi}\phi_{s}(t')}\partial_x\phi_s(t'') \rangle}{(2\pi\alpha)^2\sqrt{2\pi}}.
\end{equation}
Using  Wick's theorem for free bosons, we obtain 
\begin{eqnarray}
C&=&-\frac{i}{(2\pi\alpha)^2}\langle Te^{-i\sqrt{2\pi}\phi_s(t)}e^{i\sqrt{2\pi}\phi_{s}(t')}\rangle\nonumber\\
&&\times\left[ \langle T\phi_s(t')\partial_x\phi_s(t'') \rangle -\langle T\phi_s(t)\partial_x\phi_s(t'') \rangle \right]\nonumber\\
&=&-\frac{i}{(2\pi\alpha)^2}\frac{1}{\left[ 1+i\Lambda(t-t')\right]^2}\nonumber\\
&&\times \left[ \frac{-i/\pi\alpha}{1+i\Lambda(t'-t'')}-\frac{-i/\pi\alpha}{1+i\Lambda(t-t'')}\right]\nonumber\\
&\overset{\alpha \to 0}{=}&-\frac{2i}{(2\pi v_F)^3}\frac{1}{(t-t')(t-t'')(t'-t'')},
\label{B2}
\end{eqnarray}
for $t>t'>t''$ and $K_s=1$. Thus, the result of the bosonic calculation in Eq.~(\ref{B2}) is consistent with fermionic one in Eq.  (\ref{B1}).

\bibliography{references.bib}

\end{document}